\documentstyle[11pt,epsf]{article}

\input psfig.sty

\def\beq{\begin{eqnarray}}
\def\eeq{\end{eqnarray}}
\def\bea{\begin{eqnarray*}}
\def\eea{\end{eqnarray*}}

\def\centeron#1#2{{\setbox0=\hbox{#1}\setbox1=\hbox{#2}\ifdim
\wd1>\wd0\kern.5\wd1\kern-.5\wd0\fi
\copy0\kern-.5\wd0\kern-.5\wd1\copy1\ifdim\wd0>\wd1
\kern.5\wd0\kern-.5\wd1\fi}}
\def\ltap{\;\centeron{\raise.35ex\hbox{$<$}}{\lower.65ex\hbox{$\sim$}}\;}
\def\gtap{\;\centeron{\raise.35ex\hbox{$>$}}{\lower.65ex\hbox{$\sim$}}\;}

\def\singleandthirdspaced{\baselineskip=\normalbaselineskip\multiply
    \baselineskip by 130\divide\baselineskip by 100}

\newcommand{\newc}{\newcommand}
\newc{\qbar}{{\overline q}}
\newc{\Kahler}{K\"ahler }
\newc{\deltaGS}{\delta_{\rm GS}}
\begin{document}
\begin{titlepage}
\begin{flushright}
{\large hep-th/yymmnnn \\ SCIPP 07/17; RUNHETC-2007-24\\
}
\end{flushright}

\vskip 1.2cm

\begin{center}

{\LARGE\bf Metastable Domains of the Landscape}

\vskip 1.4cm

{\large Michael Dine$^a$, Guido Festuccia$^a$,
Alexander Morisse$^a$, and Korneel van den Broek$^b$}
\\
{\it $^a$Santa Cruz Institute for Particle Physics and
\\ Department of Physics, University of California,
    Santa Cruz CA 95064  } \\
{\it $^b$Physics Department, Rutgers University, Piscataway, New Jersey}
\vskip 4pt

\vskip 4pt

\vskip 1.5cm

\begin{abstract}
We argue that the vast majority of flux vacua with small cosmological
constant are unstable to rapid decay to a big crunch.  Exceptions
are states with large compactification
volume and supersymmetric and approximately supersymmetric states.  Neither
weak string coupling, warping, or the existence of very light
particles are, by themselves, enough
to render states reasonably metastable.  We speculate, as well, about
states which might be cosmological attractors.
\end{abstract}

\end{center}

\vskip 1.0 cm

\end{titlepage}
\setcounter{footnote}{0} \setcounter{page}{2}
\setcounter{section}{0} \setcounter{subsection}{0}
\setcounter{subsubsection}{0}

\singleandthirdspaced

\section{Introduction:  Stability in the Landscape}

There is now a widely held belief that string theory possesses a vast array
of metastable states, the {\it landscape}.  The evidence for the existence
of these states is circumstantial but, for many, compelling.\footnote{Notable
nay sayers include Tom Banks and others less vocal.}  The usual strategy is
to study a classical effective action (e.g. for IIB theories on Calabi-Yau
spaces) and to look for classically stable
stationary points.  If at the stationary point, the
system has a small coupling and large internal volume, this is strongly
suggestive that the state is sensible and metastable.

Much of the focus of landscape studies has been on states with
some degree of supersymmetry.
These are easier to study as supersymmetry provides an added degree
of theoretical control.  KKLT\cite{kklt} exhibited states in which
it appears that all moduli are fixed in a regime of weak coupling
and large compactification radius.  They (and subsequently
others\cite{banksdinegorbatov}), argued that a substantial number
of such states would exhibit dynamical supersymmetry breaking with
positive cosmological constant.
More generally,
if one considers likely mechanisms for supersymmetry breaking among the known supersymmetric
states, it is likely that a finite fraction have hierarchically small
breaking scales, as expected from conventional ideas about naturalness
\cite{branches}.

While some
degree of calculational control is valuable to theorists, however, it is
not clear why this should be important to nature.  The assumption of low
energy supersymmetry restricts the structure of the
effective action and permits inferences about
the strong coupling
and small radius regimes.  Plausible assumptions about the distribution
of the lagrangian parameters (e.g. uniform distribution of complex
parameters in the superpotential) can be checked in the
weak coupling region.  For example, one can argue that there
should be many metastable and stable states
even for small radius, and make arguments for the
distribution of supersymmetry breaking
scales and cosmological constants.\footnote{This point was first made
to one of the authors (M.D.) several years ago
by Shamit Kachru.}  But while supersymmetry provides many simplifications
and a greater degree of control, one expects that there should be more
non-supersymmetric, metastable states, possibly vastly more.
There have been some studies of the statistics of non-supersymmetric
states, both with spontaneous breaking\cite{dewolfe} and
explicit breaking\cite{dss}.  Most of these studies have involved
AdS vacua.  Attempts to study broader classes  of non-supersymmetric vacua
include those of Douglas and Denef in the case
of IIB orientifolds on Calabi-Yau spaces\cite{dd,dos}.
Their counting requires approximate supersymmetry.  To obtain finite results,
a cutoff must be imposed on the
scale of supersymmetry breaking.  Without such a cutoff, there is no control over the
calculations.  The vast majority of states are then located at the cutoff scale.
Questions such as stability against tunneling are difficult to address for such states.
A more ambitious program is that of Silverstein, who argues that there
may be various constructions which yield large numbers of
non-supersymmetric, de Sitter vacua\cite{silverstein}, and that
one may have a high degree of control.

One of the most urgent, and potentially accessible, questions in the landscape
is the origin of the gauge hierarchy.  Is it due to strong dynamics or warping,
to supersymmetry, or perhaps just anthropic selection among a vast array
of otherwise undistinguished states.   In other words, could it
be that the explanation of the hierarchy lies not in symmetries or dynamics,
but simply in the existence of an overwhelmingly
large number states which accidentally have
a small scale of weak interactions\cite{susskindhierarchy,douglashierarchy}?

The analysis of Douglas and Denef illustrates why it is hard to settle
this question.  In a typical, non-supersymmetric state, there will be
no small parameters at all, and the crutch of supersymmetry
is not available.  Unlike the supersymmetric
case, there do not seem to be any simple arguments to
give a handle on the most rudimentary statistics, much less overall
counting.

In the absence of small parameters, one question looms particularly large:
stability.  If the landscape picture has any validity, the state in which we find
ourselves, with small positive cosmological constant, sits in a large sea
of states with negative cosmological constant.  Many of these are ``close by".
Stability of any would-be state requires
that the amplitude to decay to any one of these states be very small\cite{coleman,cdl}.
Indeed, the
very notion of {\it state} requires this.  As we will see, the cutoff of Douglas
and Denef can be understood as emerging from the requirement of stability.

The significance of this last point can be understood by supposing that one has a flux landscape
of $100$ dimensions (i.e. $100$ independent
fluxes), and typical fluxes are large (say 10). Then there are of order
$10^{100}$ states.  Among these states will be states of small cosmological
constant.  Any one of these will be surrounded by many states with
negative cosmological constant.  One might
expect, for example, that there are of order $3^{100}$ within three flux units.
In order that the state be stable, it is necessary that the tunneling amplitude to any
one of these states (more precisely to the corresponding big crunch) be small.  In the
absence of a small parameter, one might imagine that there is a probability of order $1/2$
that the tunneling amplitude be zero to any one state\cite{cdl}.  So the chance that any
particular state is stable, absent any symmetries or small parameters, is of order
\beq
P_{stab} = \left ({1 \over 2} \right )^{3^{100}}.
\eeq

One might immediately object that there might be qualitative
reasons why a particular state does not decay rapidly -- or at all --
to any of its neighbors.  But this is precisely what makes this
question important: long-lived states
are likely to be special.  Optimistically, they might have features
related to phenomena we see -- or better, might hope to see, in nature.
Within the landscape, a number of classes of states have been isolated
with distinguishing features.  It is natural to ask which of these
features might contribute to stability
(In what follows,
we use ``false" vacuum to describe the
candidate metastable state; ``true" refers to any prospective decay channel):
\begin{enumerate}
\item  Weak coupling in the ``false" vacuum and/or candidate ``true" vacua.
\item  Large compactification volume
in either or both the "false" and "true" vacua.
\item  Low energy supersymmetry
\item  Light moduli in the ``false" vacuum.
\item  Warping in the ``false" vacuum.
\end{enumerate}

In this note, we investigate these possibilities.  To be concrete,
we consider mainly Type IIB theories compactified on Calabi-Yau
spaces with fluxes and an orientifold projection.
In this case, the large number of would-be metastable states
is due to a large number of possible flux choices.
There is, of course,
the risk with such specialization that our results are not
sufficiently generic.  For example, with our present
knowledge of IIB theories, it is difficult
to make statements about compactification radii (without supersymmetry), yet as we will
see, large radius is a regime (unlike weak coupling) where one
might have a realistic hope to find large metastable neighborhoods.
The constructions of \cite{silverstein} may yield vast sets of non-supersymmetric,
large volume compactifications.  We will use these,
and AdS models in IIA theory, to give some
insight into possible behaviors with volume.

In the end, of the list above, we will
argue that only the large volume and supersymmetric states are generically
stable.
The rest of this paper is organized as follows.
We first discuss some general scaling arguments for tunneling
amplitudes.  Theses arguments make clear why states
with large flux are prone to rapid decay.
  We review the argument
that supersymmetric states are stable.
We then consider the
(supersymmetric) states discussed by Giddings, Kachru
and Polchinski (GKP)\cite{gkp}.  We verify
our scaling arguments for domain wall tensions
and cosmological constants.  Because of supersymmetry,
these states are stable; we will see that this is again
consistent with simple scaling arguments.

From these exercises, we confirm that our
basic scaling arguments for domain wall tensions
and energy splittings are robust.  We then suppose that one has found a landscape of
non-supersymmetric states, and ask what features might account for stability.
We find that while weak coupling, by itself, cannot
account for metastability,  large volume -- more precisely volume scaling suitably with
flux, $N$ -- can.  Warping, in the sense discussed by GKP seems
not to lead to
stability.  The existence of approximate moduli, by itself, also does not
lead to stability.  We consider states with a small breaking of supersymmetry
(compared to the fundamental scale), and illustrate in simple models why these
are typically metastable or completely stable.

While this sort of reasoning can establish classes of states which are metastable,
it does not indicate whether one is likely to make transitions {\it into} a particular
state.  This is closely related to the questions of measures for eternal inflation
which have been widely studied recently.  While we currently have
little new to add to this discussion,
we point out that the landscape is likely to be more complicated than assumed in many
simple models of eternal inflation.  For example, a typical KKLT vacuum is likely surrounded by many
{\it AdS} states, both supersymmetric and not.
Whether one can neatly
transition into the KKLT minimum seems a serious question.  We speculate
that states with (discrete) symmetries, though rare,
might be attractors in cosmological evolution.

In the conclusions, we indulge in conjecture.  Our results, we note, hardly prove
that low energy supersymmetry is a feature of the landscape, but they suggest, in ways
we explain, that it might be.  They suggest, alternatively, what is required to establish
the existence of a vast set of non-supersymmetric states in the landscape.


\section{Scaling Arguments for Non-Supersymmetric States}
\label{nonsusy}

In this section, we ask how we might expect potentials,
domain wall tensions, and tunneling
amplitudes to scale in the limit of large fluxes, in non-supersymmetric
states.
In order that
in these hypothetical states there be some validity to a semiclassical
analysis in a ten dimensional
effective field theory, we will suppose that
the compactification volume, $V$, is large.  We will also
assume, when necessary, that couplings are small.
We suppose that we have many three-form fluxes ($b$), with typical values
of order $N$.  The potential is quadratic in $N/V$ (in the Einstein
frame, i.e. in four dimensional Planck units).

We are considering a Type IIB landscape, with various RR and NS 3 form fluxes,
$N_i$ and $K_i$.  We will think of the fluxes as very large, $N_i \sim N$,
and $K_i \sim K$, for some large $N$ and $K$.
We wish to compare neighboring states in the landscape, i.e. states in which
one flux, say, $N_i$, changes by one, or perhaps a few fluxes change by a few.
Take the first case.  For simplicity, suppose first that $K \sim N$.  In this case,
there is no particular reason for a semiclassical approximation to be valid,
but we will use the classical formulas for the potential in order to get some
feeling for how amplitudes might scale with $N$ in a ``typical" state.  Later,
we will adjust the fluxes so that the string coupling is weak.

When all fluxes are comparable,
because the potential is homogeneous in $N$,
the changes in the moduli fields are of order $1/N$, and the change in the potential
is of order $N$ for small changes ($\Delta N \ll N$) in flux.

In the decays of interest
to us, both the four dimensional
fields and the fluxes change, and General Relativity
plays an important role..
But first it is worthwhile to review some aspects of tunneling in ordinary
field theory, without gravity.
For our discussion
it is important to recognize
that even though the barrier may be quite high, if the neighboring well
is very deep and the field
excursions are not too large, the tunneling amplitude is not necessarily small.

Consider a theory of a scalar field $\phi$,
in $4$ dimensions, with
a potential of the form:
\beq
V(\phi) = {N^2 \over V^2} f(\phi).
\eeq
Here $N$ is supposed to be large, as in our problem above, and $f(\phi)$ is a function
with two minima.
Let's first ask about the range of validity of the semiclassical analysis in
the two would-be minima.  Assuming a canonical kinetic term, corrections to the
kinetic term, at one loop, behave as $N^2 \over V^2$
(the vertices each give a factor of $N^2 \over V^2$, and there is a factor of
$1/m_{\phi}^2$ from the integral).  So the perturbative analysis would seem to be valid
if $V \gg N$.  We will see that stability seems to give a somewhat stronger condition.

Turning to tunneling, ignoring gravity,
the bounce action behaves
as $V^2/N^3$.  This follows from simple
scaling arguments on the terms
in the action.  But it can be seen by considering the
standard thin wall analysis, which will be valid in
the case that the two minima of $f$ are nearly
degenerate, differing by $\Delta E$.  Then the bubble tension is
given by ($\Delta \phi \sim 1/N$)
\beq
T =  \int^{\Delta \phi} d\phi \sqrt{2 V(\phi)} \sim {1 \over V},
\eeq
so
the standard thin wall analysis\cite{coleman} gives
\beq
S_b = C {T^4 \over \Delta E^3} = A V^2/N^3,
\eeq
where $C$ and $A$ are numerical constants.  So unless the volume is large (of order
$N^{3/2}$ in fundamental units), the bounce action is small and
the tunneling amplitude is of order one.  This same scaling
can readily be shown to hold for more general functions $f$.

Typically, gravitational corrections will be large.  From the
Coleman-DeLuccia analysis, in the case of tunneling, say, from flat or AdS space to AdS
space, one learns that gravitational corrections are important when the
radius of the would-be bubble is comparable to the AdS radius.  In our example above,
the radius of the bubble wall is:
\beq
R_b = {T \over \Delta E} \sim V/N
\eeq
while the AdS radius is of order $V/N$.

On the other hand, for the case of interest to us in the landscape, the initial state,
by assumption, has
zero or nearly zero cosmological constant.  So,
\beq
R_{AdS} \sim {V \over \sqrt{N}}
\eeq
and one does not expect gravitational corrections to be important.
More generally, one does not expect a suppression of decays from states
with large positive to negative cosmological constant.

Of course, by definition, if the bounce action is not large, the semiclassical
calculation is not reliable.  We take this result as evidence that suppressing
tunneling requires $N^3 \ll V^2$.  Note that this constraint is more severe
than the naive perturbative condition, $V \gg N$.

When we consider flux vacua,
we will be interested in the Coleman or Coleman-DeLuccia analysis in cases
where the bubble can be thought of as a brane ($M5$ or $D5$) wrapped
on some internal manifold.  Such branes can be thought of as thin-walled
bubbles.  To make the comparison,
it is useful to reformulate the conventional field theory analysis in terms
of a collective coordinate.   If the bounce is described
by a function $\phi_{cl}(r-R)$ (spherically
symmetric), we can introduce a collective coordinate, $R(t)$,
and obtain an action for $R$ by writing:
\beq
\phi(r) = \phi_{cl}(r-R(t)):
\eeq
\beq
S = \int dt \left (T R^2 \dot R^2 - V(R) \right )
\eeq
with
\beq
V(R) = T R^2 -\Delta E R^3.
\eeq

The bounce action follows from an ordinary WKB calculation
with this action.\footnote{In terms of our earlier remark about quantum tunneling, it is the
unusual form of the kinetic term for $R$ and the
pressure term which account for the enhanced tunneling rate.}  For the case
of wrapped branes, this is the structure of the appropriate
Born-Infeld action.

The estimates in this section
suggest that the tunneling amplitude quickly becomes of order one as $N$ increases,
if $V$ and the couplings are fixed.  But one
might worry that this field theory analysis is not applicable to the
case of interest, where transitions are accompanied by changes of flux and
emission of extended objects like branes.  In the next section, we will
see that precisely these scalings of tensions
and cosmological constants occur in well-studied string theory
examples.

\section{A Prototype:  GKP}

A useful model for stabilized moduli and domain wall tensions is provided by
the work of Giddings, Kachru and Polchinski.  This is a IIB orientifold model,
with compactification on a Calabi-Yau space at a point in the
(approximate) moduli space near a conifold singularity.  In this model,
the moduli are $\tau$, $\rho$ and $z, z_i$.  $\tau$ is the IIB string coupling;
$\rho$ describes the overall size of the compact manifold,
\beq
\rho = R^4 M_{10}^4
\eeq
(where $M_{10}$ is the ten-dimensional Planck scale), and $z$ is
a complex structure modulus which describes the
deformation of the conifold.  The $z_i$ represent other complex structure moduli.
$\tau$, $z_i$ and $z$ are fixed by fluxes; $\rho$ is undetermined in a semiclassical
treatment (the model has a no-scale structure).  It will be useful to consider
an effective field theory without $\rho$, as well, in which supersymmetry
is unbroken.

For large $\rho$, the theory turns out to be approximately supersymmetric.  Following
GKP, we will consider light fields $z$, $\tau$ and $\rho$.
Their dynamics
can be described by a superpotential and Kahler potential.  The superpotential
takes the form:
\beq
W = M {\cal G}(z) - K \tau z - K^\prime \tau h(z)
\eeq
where $M$ is the RR three-form flux along the $A$ cycle associated with the
conifold;
$K$ is the NS-NS three form flux along the corresponding $B$ cycle; and
$K^\prime$ is the flux along some other cycle, $B^\prime$.
The function ${\cal G}$ has the form:
\beq
{\cal G}(z) = {z \over 2 \pi i} \ln(z) + {\cal G}(0) + f(z)
\eeq
where $f(z)$ is a holomorphic function of $z$ which vanishes as $z \rightarrow 0$.
The Kahler potential
\beq
K = -3 \ln(\rho - \rho^*) - \ln(\tau - \tau^*) + f(z,z^*)
\eeq
where $f(z,z^*)$ is a finite function of $z$ which tends to a constant as $z \rightarrow 0$.

For suitable choices
of flux, the equations for supersymmetric stationary points of
$z$ and $\tau$ have solutions with $z$ small and $\tau$ large:
\beq
z= {\rm exp}({{2 \pi K \, {\rm Im}~\tau \over M}}) ~~~~~
{\rm Im}\tau = -{M {\cal G}(0) \over K^\prime h(0)}.
\label{ztau}
\eeq
As noted above, at these points, $\rho$ is undetermined and the potential, classically,
vanishes.  For large $\rho$, the gravitino mass
is small and one can argue that the computation is self-consistent.
The masses of the lightest Kaluza-Klein modes are of order
\beq m_{KK}^2 \sim {1 \over \rho^2},\eeq
where this, which is expressed
in four dimensional Planck units,
applies in the limit of large $\rho$ and moderate $z$.

The masses of the gravitino, $m_{3/2}$, is of order:
\beq
m_{3/2}^2 \sim \vert M {\cal G}(0) \vert^2 \rho^{-3} g_s,
\eeq
justifying the use of a the supersymmetric lagrangian, for large $\rho$ (and/or
small $z$).  The masses of $z$ and $\tau$ are of order:
\beq
m_\tau^2 \sim \vert M{\cal G}(0) \vert^2 {g_s \over \rho^3}~~~~m_z^2
\sim M^2{g_s\over \rho^3 z^2}.
\eeq
So the inclusion of $\tau$ in the low energy effective lagrangian is sensible; for fixed
$\rho$, however, $z$ becomes massive as $z \rightarrow 0$, and a more careful
analysis seems required.  We will proceed without worrying about this potential
subtlety, as we will be generally interested in rough scalings of tunneling
amplitudes, in any case.

\subsection{Tensions and Cosmological Constants}

The vacua in this leading approximation are all degenerate (zero
cosmological constant), so
this model is not useful for discussing tunneling amplitudes.  As a toy model,
we can consider a theory without
$\rho$, with minima which are supersymmetric and AdS.  However,
because of supersymmetry, there is still no
tunneling between the vacua.  The model is useful, however, for studying how domain
wall tensions and vacuum energies (cosmological constants) scale with flux, and also
how warping effects these quantities.  Without $\rho$ (or with $\rho$ fixed),
the domain wall tension between vacua of different flux
is\cite{domainwalltensions}:
\beq
T =2 \Delta \vert {e^{K/2} W}\vert.
\label{generaltension}
\eeq
We can consider various types of transitions.  For transitions with $\Delta M= \pm 1$,
the tension is given by
\beq
T \approx 2 \left \vert {\cal G}(0) e^{K/2} + e^{K/2}( K^\prime h(0)
- {1 \over 2 \tau} [M {\cal G}(0) + \tau K^\prime(0) ])\Delta \tau \right \vert
\label{deltamone}
\eeq
$$~~~~\sim 2 \vert{\cal G}(0) \vert {\rho^{-3/2} g_s^{1/2}}.$$
Where we used that at the minima $\tau=-\bar{\tau}$ and neglected contributions of order $z$.
This is in accord with the general scaling arguments of section \ref{nonsusy}; the
tension is of order one, if all fluxes are scaled uniformly.
In addition, in the theory absent $\rho$, the change in the cosmological constant
is of order $M$, again in accord with our general scaling arguments.

For those with $\Delta K^\prime = \pm 1$, the tension is of order:
\beq T \sim {M/K^\prime} {\rho^{-3/2} g_s^{1/2}}.\eeq
Again, if all scales become uniformly large, this is in accord with our earlier
arguments.  Interestingly, however, for small string coupling, this is enhanced
relative to the $\Delta M=1$ transitions.  The change in the cosmological constant
is of order
\beq
M^2/K^\prime \rho^{-3} g_s.
\label{deltakprimeone}
\eeq
Finally, changes in $K$ are associated with domain walls with tension suppressed by $z$,
and with changes in cosmological constant similarly suppressed.

\subsection{GKP As a Prototype for Non-Supersymmetric Scalings}

Because of the supersymmetry of the GKP solution, without $\rho$, or the degeneracy
with $\rho$, semiclassically there is no tunneling.  But we have seen that
the domain wall tension and energy splittings behave as expected for theories without
supersymmetry, so
we can use the GKP solution as a model for
transitions among non-supersymmetric
states.   We would then expect that tunneling
amplitudes would behave roughly as $e^{-S_b}$, where $S_b \sim T^4/\Delta E^3$.
So, for example, for the
for
transitions with $\Delta M=1$, we would have, for the bounce action:
\beq
S_b = {e^{-K} \over M^3} \sim {V^2 \tau  \over M^3}\sim {V^2 \over M^2 K^\prime}.
\label{deltamsb}
\eeq
Here we have used the flux dependence of $\tau$, eqn. \ref{ztau}
This flux and volume behavior is as we anticipated in section \ref{nonsusy}, in the
sense that it involves three powers of flux in the denominator, though one of these
factors is the small (NS-NS) flux.

In any case, this model is compatible with our naive estimates:  obtaining a large set of
metastable
non-supersymmetric compactifications would seem to require that the volume
scale as a power of the flux.  Following our
discussion in section \ref{nonsusy}, we would expect that, while
in general, gravitational corrections are important and might
suppress the decay amplitude in some cases,
this is not the case if the initial state has small cosmological constant.

This has a close parallel to the field theory discussion along
the lines of \cite{coleman,cdl}.
For the case of a change
of one unit of RR flux, the expanding
bubble can be thought of as a wrapped
$D5$ brane.  Three of the directions along the brane are wrapped on a three cycle; the
remaining two dimensions correspond to the bubble wall.  Of the four collective coordinates
of the brane, three are located at a point on the internal manifold; the remainder is the
coordinate, $R(t)$, describing the wall surface.  The tension of the D5 brane is just
the tension we identified before:
\beq
T = {1 \over g_s} M_s^6 (R^3 M_{10}^3) M_{10}^{-3}
\eeq
$$~~~~={1 \over g_s} \rho^{-3/2} M_p^3$$
which is what we found from the heuristic field-theoretic argument.

We need, also, to worry about conservation of $D3$ brane charge.  In general, transitions
involving changes in flux will involve emission of $D3$ branes, as well.
We won't consider this problem in detail.  However, for special cases
(those for which
$
\int H \wedge F$
vanishes), there need be no $D3$ brane emission.  In cases where there is brane
emission, the energetics are the same as suggested by the field theory arguments.
For a change in $K$ of order one, one needs a change in the $D3$ brane density
of order $M$.  Wrapped $D3$ branes have tension (mass) of order $1/g_s M_p$ (independent
of $\rho$), so the energy density is of the same order as the change in energy due to
the change in flux.  So we expect that our estimates of tunneling rates above are still correct.

\section{Tunneling From Approximately Supersymmetric Vacua}

From the point of
view of stability, supersymmetric vacua are special.
According to quite general arguments, they are stable.
This is perhaps surprising, since in supergravity it is perfectly possible for a
lagrangian to exhibit a supersymmetric
state (say with vanishing cosmological constant) and a non-supersymmetric state with
large, negative cosmological constant.  In such a case, one does not expect, for example,
a BPS domain wall.

The essential point was made by Deser and Teitelboim\cite{deser}, Witten\cite{wittenpositive},
Hull\cite{hullpositive} and others long ago.  They noted
that in a classical supergravity theory in an asymptotically flat space, one can
define not only a total energy and momentum, but also global supercharges.  These obey the
standard supersymmetry algebra, so, just as is familiar in global supersymmetry, the
energy of any configuration can be shown to be greater than or equal to zero.  (This
is a special case of the positive energy theorem.)

The stability of exactly supersymmetric states may or may not be of interest, but clearly
the stability of {\it approximately} supersymmetric states is of great potential
importance.
Suppose the scale of supersymmetric breaking, $F$, is small compared to the Planck
scale and other possible scales of interest (string scale, compactification scale).
Then the decay probability behaves as
\beq
\Gamma \sim e^{-1/\vert F \vert^2}
\eeq
or vanishes.  It is only non-vanishing if susy breaking in the AdS state is comparable
or smaller than that in the approximately flat space state.

\subsection{Models}

These features of decays of (nearly) supersymmetric states can be illustrated
with simple models.  Consider, first, a theory of a single scalar field, $\phi$,
with superpotential:
\beq
W= {1 \over 2} M\phi^2  -{1 \over 3} \gamma \phi^3.
\eeq
Before coupling to gravity, this theory has supersymmetric minima at
\beq
\phi_0=0;~\phi_0 = {M \over \gamma}.
\eeq
These can be joined by a domain wall, with tension:
\beq
T = 2 \Delta W.
\eeq
Now couple the system to (super)gravity, with $M \ll M_p$.  In this case, the domain
wall tension is approximately unchanged, but there is a splitting between the states,
\beq
\Delta E= 3 {\vert \Delta W \vert^2 \over M_p^2}
\eeq
$\Delta E$ is small compared to the scales in the superpotential, so a thin wall approximation
is appropriate.  The bubble radius is, again,
\beq
R_b = {3 T \over  \Delta E}
\eeq
As explained by Coleman and DeLuccia, the decay amplitude vanishes if
\beq
{R_b \over \Lambda} =2
\eeq
and precisely this condition is satisfied in this model.

Now let's add supersymmetry breaking to the mix.
This can be done by adding an additional chiral field, $Z$,
and taking for the superpotential:
\beq
W = {M \over 2} \phi^2 - {\gamma \over 3} \phi^3 + Z\mu^2 + W_0.
\eeq
The $Z$ field can be stabilized in both vacua by adding a term $Z^\dagger Z Z^\dagger
Z$ to the Kahler potential.  $W_0$ is chosen so that the $\phi=0$ state
has zero cosmological constant:
\beq
3 \vert W_0 \vert^2 = \vert \mu^4 \vert .
\eeq
Note that the phase of $W_0$ is not fixed by this condition.

For this system, the bubble wall tension is approximately
as it was in the previous case,
but the energy shift is different.  Calling the original shift $\Delta E_0$, the
shift is larger or smaller by an amount of order $F M^3/(\gamma^2M_p)$, depending on the sign
of $W_0$.  In the latter case, the amplitude vanishes; in the former, it is of order
\beq
\Gamma \approx e^{-{6 \pi^2} M_p^4/\vert F \vert^2}.
\eeq

Note that even for moderately small $F$ (in whatever are the appropriate units) and in the
presence of an exponentially large number of decay channels, the decay
rate is extremely small.  We might expect that $\vert F \vert \sim
10^{-3}~M_p^2$ would more than adequately
suppress the decay rate.
The implications of this observation for a possible {\it prediction}
of low energy supersymmetry will be discussed in the conclusions.

\section{Large volume, Weak Coupling, Light Moduli and Warping}

From our studies of the GKP model, we can already see that weak string coupling,
by itself, does not insure stability.  However, a combination of weak coupling
and large volume does.  What is required is that there be a large number of states
whose volume scales with a power ($3/2$) of the flux.  One suspects, more
generally, that whatever is the parameter(s) which account for the exponentially
large number of states, the volume must scale with a power of this parameter.

In the IIB constructions which have been studied, the fixing of the volume is not well
understood (except in special cases which have exact
or approximate low energy supersymmetry).
In IIA theories, however, candidate vacua have been identified with all moduli
fixed\cite{dewolfe,dss}.  In these cases, one has an infinite sequence of
AdS vacua, with or without supersymmetry, with progressively larger volume and
smaller cosmological constant.  We can again take these as a model, supposing
that there exists a set of dS vacua with similar scalings, and ask how
the tunneling amplitudes would behave.

Without reviewing the IIA models in detail, we note that the important large number
in these constructions is the four-form flux, which we will refer to generically
as $N$.  As in the IIA case, the action scales as $N^2$.  At the minimum, one
finds that the volume, $v$, and the dilaton, $e^{\phi}$, scale as
\beq
v \sim N^{3/2} ~~~~ e^{\phi} \sim N^{-3/4}
\eeq
and the cosmological constant behaves as $N^{-9/2}$.  As a result,
the tension of the bubble walls and the energy
splittings between states behave as
\beq
T \sim N^{-13/4} ~~~~ \Delta E \sim N^{-11/2}
\eeq
giving a result for the bounce action which grows rapidly with $N$:
\beq
S_b \sim N^{3.5}.
\eeq

Another model for scaling with volume is provided by the proposal of Silverstein
and Saltzman to compactify string theory on products of Riemann surfaces\cite{silverstein}.
Again, without reviewing the details of the model, there are various
numbers, such as $5$-form flux, $q_5$, which can become large, accounting
for the large number of states, many of which
are believed to be de Sitter.  The volume, in this case,
scales as $q_5^{3}$, while the vacuum energy scales as $V^{-4/3}$.  So again
the bounce action grows as a power of the large parameter.

Both of these models suggest that if there do exist large sets of de Sitter
vacua with growing volume and decreasing cosmological constant, they are likely
to be highly metastable.

Warping, on the other hand, does not
seem, by itself, to lead to suppression of tunneling rates, as we see
from the GKP model.  One might have expected this in any situation where
there are large numbers of fluxes.  The GKP construction suggests that warping
is obtained by tuning some set of fluxes on cycles associated with the warp
region; changing far away fluxes, then, may be possible without
spoiling this feature. Formulas \ref{deltamone}-\ref{deltakprimeone}
exhibit no singular dependence on $z$.

\section{Tunneling from states with Light Moduli}

Supersymmetric vacua with moduli are very familiar.
Much of the recent focus on flux vacua is motivated by
the observation that these have few or no light moduli.
Still, we might speculate that there exist classes of
non-supersymmetric vacua with comparatively light pseudomoduli.
Those with small cosmological constant would likely have neighbors
with negative cosmological constant, and no moduli at all.

One can ask whether the presence of light moduli
would somehow suppress tunneling.  In the standard treatment (we will
ignore the effects of gravity in this section) one needs to study an analog
problem, the motion of a particle in a potential, with boundary conditions
that in the far future, the system settle into the false minimum of the potential.
One might hope that the light field would only slowly settle into its minimum,
giving rise to a large bounce action.  This turns out not to be the case, in general.

In the transition, one expects a significant rearrangement of the degrees of freedom.
The final state in the transition likely will have no light fields, or at least different
numbers of them.  High energy string states in one vacuum might be relatively light
states in another.  To develop some intuition, we consider a field theory with
two fields, $X$ and $\phi$. $\phi$ is light in the ``false" vacuum but
heavy in the ``true" vacuum; $X$ is heavy in both.  For the potential we take:
\beq
V = {1 \over 2} \mu^2 \phi^2 + {M^2 \over 2} X^2
- {1 \over 4} X^4 + {\Gamma \over 6} X^6 + \epsilon X^2 (\phi -\phi_0)^2.
\eeq

The idea here is that $\mu$ is extremely small compared to $M$, $\Gamma$, which define
mass scales of the same order.  To permit simple calculations, we can take $\epsilon$
to be a small number.
$\Gamma$ is chosen so the vacuum with $X \ne 0$ has lower energy than the $X=0$
vacuum.  Again, to allow simple approximations, we
can tune $\Gamma$ so that the energy difference
between the $X=0$ vacuum and the $X \ne 0$ vacuum is, say, of order $\epsilon M^4$.

For small $\epsilon$, we can consider first the dynamics of $X$ by itself.  We will
also ignore gravity at first.
This is then a standard thin wall tunneling problem.
The bounce has a size of order $r_0 = t_0 =1/(\epsilon M)$, where $r_0$ is meant to denote
the bubble radius and $t_0$ denotes the typical time in the analog particle problem.
In the particle analogy with inverted
potential, $X$ starts extremely close to the top of the hill (the true
vacuum),
\beq
X - X_0 \sim M e^{-M t_0}.
\eeq
It then rolls quickly to the false vacuum.  As it approaches the false vacuum
($X=0$), it behaves, again, as
\beq
X \sim M e^{-M(t-t_0)}.
\eeq
For sufficiently large time ( $M (t-t_0) \sim \ln({\epsilon M^2\over \mu^2} )$),
$\epsilon X^2 \approx \mu^2$.  Until that time, the minimum
of the $\phi$ potential lies at $\phi_0$.  In the inverted problem, $\phi_0$ then
rolls away from
that point (towards the origin).

Note, however, that this time is much smaller than $\mu^{-1}$ (it is also
smaller than $r_0$).  So $\phi$ satisfies the
equation:
\beq
\ddot \phi + {3 \over t} \dot \phi =0.
\eeq
This has solutions $\phi=1/t^2+ {\rm constant}$.  By tuning the initial conditions,
one can arrange that $\phi < \mu$ in a time much less than $\mu^{-1}$; for such motions,
there is no enhancement of the bounce action.

\section{Implications and Speculations}

From this survey, we have concluded that generic, metastable states, are likely
to satisfy special conditions.  We have identified two possibilities:
\begin{enumerate}
\item  (Approximate) Supersymmetry at scales well below the fundamental scale.
\item  Compactification radii much larger than the string
scale
\end{enumerate}
and ruled out several others.

The strongest indication for the existence of large numbers of large volume, dS states
comes from the work of \cite{silverstein}, but the
work of \cite{dd,dewolfe} is also suggestive.
For many of these constructions, there are good reasons for skepticism (see, e.g.,
\cite{bankskorneel}, though it should be noted that there are also
reasons to be skeptical of supersymmetric constructions).
Even if such states exist, it is conceivable that many
are in some sense uninteresting, since coupling constants become small
in the large volume limit.  If, in the end, the explanation of hierarchy is the
existence of a vast array of non-supersymmetric states (with large volume
or some other feature which accounts for metastability), the question
will be:  are these states distinguished in some other way?  Otherwise,
observations of phenomena at accelerator energies will provide us with no
information about physics at extremely high energies.

One might speculate that there
do not exist such large sequences of large volume compactifications,
no some other generic vast class of non-supersymmetric metastable dS states.
Stability might favor low energy supersymmetry.  It is then interesting to ask
whether among these states, there might be further selection effects.

\subsection{R Symmetric Points as Cosmological Attractors}

Having established that some set of states are metastable, it is natural to ask
whether the universe might find its way into them.  To address this issue, the first
task is to survey the neighborhood of some particular state (or class of states)
of interest.  Let's begin with the KKLT vacua.  It seems likely that these are surrounded
by AdS vacua, largely non-supersymmetric.
From our arguments above,
the KKLT vacua are likely stable against
decays to these states.
Because of these many AdS states, transitions {\it into} the KKLT vacua might be difficult.

A potentially interesting set of states are those which exhibit discrete R symmetries.
Such states are likely rare in the landscape\cite{dinesun}, but their neighborhoods
may be more interesting.  In general, one finds such states by setting to zero all
fluxes which transform non-trivially under the candidate symmetry.  Typically, this is
a substantial fraction of the fluxes -- 2/3 or more.

Arguably this is what one expects for symmetric states:  they are special
and thus rare.  But the idea that cosmological considerations (e.g. high
temperatures) can favor symmetric states is familiar. For the flux lattice,
things cannot be so simple;
we are concerned about discrete transitions.  But it seems possible that such symmetric
states might be cosmological attractors.
Consider the neighborhood of the $R$ symmetric states, i.e. states for which
the symmetry preserving fluxes, $N_i$, are large, while those which
break the symmetry are much smaller, say $n_\alpha$.
Similarly, there are
fields, $Z_i$, which transform like $W$ at the $R$ symmetric point, and fields, $\phi_a$,
which transform differently.  At the $R$ symmetric point, there is a superpotential:
\beq
W = Z_i f^i(\phi_a,Z_i)
\eeq
The condition for unbroken supersymmetry is:
\beq
Z_i =0;~~f_i(\phi_a,0) =0.
\eeq
Provided there are more $\phi_a$ type fields then $Z_i$ fields, these equations,
generically, possess a moduli space of solutions.

Now treat $N_i$ and $n_\alpha$ as sufficiently large
that they can be thought of as continuous, with $\vert N_i \vert \gg \vert
n_\alpha \vert$, for all $i$ and $\alpha$.  As we
turn on small $n_\alpha$ we can study an effective lagrangian for
the light moduli of the symmetric
vacuum.   There is no reason to think that the structure
of stationary points of the effective lagrangian for this potential is different
than the general (non-R) case.  So there will typically be solutions with broken
supersymmetry, and positive or negative cosmological constant\cite{dd}.
Introducing polar coordinates,
$\vec n = (n,\theta_1,\theta_2,\dots)$.  The cosmological constant, as a function
of $\vec n$, then has the structure:
\beq
\Lambda(\vec n) = n^2 v(\theta_i).
\eeq
It is then plausible that there are finite elements of ``solid angle" with either sign of the
cosmological constant, and in particular finite regions with positive
cosmological constant and energy tending
to zero as $n \rightarrow 0$.  For these regions, the symmetric state
may function as a {\it cosmological attractor}.  Since most ``jumps"
will be very rapid, it is perhaps appropriate to think of $\vec n$ as
a collective coordinate, and the motion in this
space as reasonably smooth.  Note that in the last few transitions,
the continuous approximation for $n$ will break down, and there may be AdS
states close to the symmetric point, which may have catastrophic consequences for a particular decay chain.  But it is at least plausible that finite domains of the landscape are attracted to R symmetric vacua.

\subsection{Implications of Stability}

Of cosmological issues within the landscape, metastability is the simplest criterion
which states must satisfy.  Of generic features of landscape states, we have seen that
only large volume, or approximate supersymmetry, seem to result in some degree of
metastability.   It is conceivable that we could settle the question of whether
vast arrays of large volume, non-supersymmetric states exist in an underlying
theory of gravity.  As we have noted, if they do, and they are otherwise
undistinguished, it is unclear how one might
imagine developing a string phenomenology. Not only would we fail to make
predictions, e.g. for LHC physics, but we would not know how to
interpret LHC outcomes.  If not, however, low energy supersymmetry
would seem a prediction of string theory.

We should stress, of course, that stability, by itself, does not imply {\it weak scale}
supersymmetry.  A quite high scale of supersymmetry
breaking will insure an adequate level of stability.
However, within the supersymmetric branches of the
landscape, the scale of supersymmetry breaking is likely flat on a log scale\cite{branches}.
So, just as expected from conventional naturalness arguments, light Higgs should
arise with low supersymmetry breaking scale.

We have gone further, examining the neighborhood of the KKLT vacua, as well
as sets of vacua with approximate $R$ symmetries, suggesting reasons why
cosmology might favor the later.  These remarks are on far shakier ground,
but are worthy of further study.

\noindent
{\bf Acknowledgements:}
We thank A. Aguirre, N. Arkani-Hamed, T. Banks, E. Silverstein and S. Kachru for conversations.  The argument about
light moduli was developed in discussions with T. Volansky.
The work of M.D., G.F. and A.M. was supported in part by the U.S.
Department of Energy.  Part of this work was done at the Kavli Institute for Theoretical
Physics and at the Institute for Advanced Study; M.D. thanks both institutions for their
hospitality and support.  M.D. received support from the J.S. Guggenheim Foundation for
part of this work.  KvdB would like to thank SCIPP, where part of this work was done,
for their hospitality. This work is supported in part by the Rutgers Department of Physics and
Astronomy and DOE grant number DE-FG02-96ER40949

\end{document}